\documentclass[twocolumn,preprintnumbers,amsmath,amssymb]{revtex4}

\usepackage{graphicx}
\usepackage{dcolumn}
\usepackage{bm}
\usepackage{epstopdf}
\usepackage{amsmath}

\begin{document}

\title{Universality of Zipf's Law}

\author{Bernat Corominas-Murtra$^1$ and  Ricard V. Sol\'e$^{1,2,3}$}


\begin{abstract}
Zipf's  law is  the  most common  statistical distribution  displaying
scaling behavior.   Cities, populations or firms are  just examples of
this  seemingly universal  law.  Although  many different  models have
been proposed,  no general theoretical  explanation has been  shown to
exist for its universality.  Here we show that Zipf's law is, in fact,
an inevitable outcome  of a very general class  of stochastic systems.
Borrowing concepts from Algorithmic Information Theory, our derivation
is based on  the properties of the symbolic  sequence obtained through
successive  observations over  a  system with  an  ubounded number  of
possible states.   Specifically, we assume that the  complexity of the
description of the system provided  by the sequence of observations is
the one expected for a system evolving to a stable state between order
and  disorder. This  result  is obtained  from  a small  set of  mild,
physically relevant assumptions.  The general nature of our derivation
and its model-free  basis would explain the ubiquity of  such a law in
real systems.
\end{abstract}

\affiliation{  $^1$  ICREA-Complex  Systems  Lab,  Universitat  Pompeu
  Fabra, Parc de Recerca Biom\`edica de Barcelona (PRBB) Dr.  Aiguader
  80, 08003 Barcelona, Spain\\  $^2$Santa Fe Institute, 1399 Hyde Park
  Road,   New  Mexico   87501,   USA\\  $^3$   Institut  de   Biologia
  Evolutiva. CSIC-UPF. Passeig Maritim de la Barceloneta, 37-49, 08003
  Barcelona, Spain}

\maketitle

\section{Introduction}

Scaling  laws  are  common  in  both natural  and  artificial  systems
\cite{Stanley:2000}.  Their  ubiquity and  universality is one  of the
fundamental   issues   in   statistical  physics   \cite{Stanley:1999,
  Newman:2005,  Sole:2001}.  One  of  the most  prominent examples  of
power   law   behavior   is   the   so   called   {\em   Zipf's   law}
\cite{Zipf:1949,Auerbach:1913,  Gabaix:1999}.  It  was  popularized by
the  linguist G.   K.  Zipf,  who observed  that it  accounts  for the
frequency   of    words   within   written    texts   \cite{Zipf:1949,
  Ferrer:2002}. But this law is extremely common, \cite{LiWeb}
and has  been found in the  distribution of populations  in city sizes
\cite{Zipf:1949,  Simon:1955,  Makse:1995,  Krugman:1996,  Blank:2000,
  Decker:2007}, firm sizes in industrial countries \cite{Axtell:2001},
market      fluctuations     \cite{Gabaix:2003},      money     income
\cite{Pareto:1896, Okuyama:1999}, Internet file sizes \cite{Reed:2002}
or family names \cite{Zanette:2001}.  For instance, if we rank all the
cities  in a  country from  the largest  (in population  size)  to the
smallest, Zipf's law states that the probability $p(s_i)$ that a given
individual lives in the $i$-th most populated city ($i=1,...,n$) falls
off as
\begin{equation}
p(s_i) =\frac{1}{Z}i^{-\gamma},
\end{equation}
with the exponent, $\gamma \approx 1$, and being $Z$ the normalization
constant, i.e.,
\begin{equation}
Z=\left(\sum_{i\leq n}i^{-\gamma}\right).
\label{Normal}
\end{equation}
Although  systems   exhibiting  Zipf's-like  statistics   are  clearly
different in their constituent units, the nature of their interactions
and  intrinsic   structure,  most  of  them  share   a  few  essential
commonalities. One  is that they are stochastic,  far from equilibrium
systems changing in time, under mechanisms that prevent them to become
homogeneous.   Within the  context  of economic  change, for  example,
wider varieties of goods and attraction for people are fueled by large
developed areas.  Increasing returns drive further growth and feedback
between  economy  and  city  sizes  \cite{Krugman:1996b,  Arthur:1994,
  Manrubia:1999}.   Moreover,  the presence  of  a  scaling law  seems
fairly  robust through  time:  in spite  of  widespread political  and
social changes,  the statistical behavior  of words in  written texts,
cities or firms  has remained the same over  decades or even centuries
\cite{Zipf:1949,      Axtell:2001,      Rozman:1990,      Gabaix:1999,
  Manrubia:1999}.   Such  robustness  is  remarkable,  given  that  it
indicates  a  large  insensitivity  to multiple  sources  of  external
perturbation.  In spite of their disparate nature, all seem to rapidly
achieve the Zipf's law regime and remain there.

To account  for the  emergence and robustness  of Zipf's  law, several
mechanisms  have  been  proposed, including  auto-catalytic  processes
\cite{Levy:1996,   Biham:1999,    Huang:2000},   extinction   dynamics
\cite{Sole:1996,   Newmann:2003b},  intermittency  \cite{Zanette:1997,
  Manrubia:1998},       coherent       noise       \cite{Newman:1996},
coagulation-fragmentation  processes  \cite{White:1982,  Family:1985},
self-organized  criticality  \cite{Bak:1987}, communicative  conflicts
\cite{Harremoes:2001,    Ferrer:2003},   random    {\em   typewriting}
\cite{Li:1992,       Perline:1996},       multiplicative      dynamics
\cite{Montroll:1982, Kawamura:2002} or stochastic processes in systems
with    interacting   units    with    complex   internal    structure
\cite{Amaral:1999}.  The diverse  character of such mechanisms sharing
a common scaling exponent  strongly points towards the hypothesis that
some   fundamental  property  (beyond   a  given   specific  dynamical
mechanism)  is at work.   Such a  universal trend  asks for  a generic
explanation, which should avoid the use of a particular set of rules.

We address the problem  from a very general, mechanism-free viewpoint;
by studying  the statistical properties of the  sequence of successive
observations over the system.  More precisely, our observations can be
understood as a sequence of  symbols of a given alphabet (depending on
the  nature of  the system)  following some  probability distribution.
The elements of  this alphabet can be coded in  some way -for example,
bits.  From  this conceptual starting  point, we borrow  concepts from
algorithmic information theory (AIT) and propose a characterization of
a wide family of stochastic systems, to which those systems displaying
Zipf's  law would  belong.   Such a  characterization imposes  special
features  on the  behavior of  the entropy,  whose study  leads  us to
conclude that,  under generic mathematical assumptions,  Zipf's law is
the only solution.

The paper is organized as  follows: In section II we briefly introduce
the concept of stochastic object  as defined within the context of AIT
and how it helps to understand our problem. In section III we find the
asymptotic   solutions    of   the   equations    derived   from   the
characterization  provided in  section II.   Section IV  discusses the
relevance of the obtained results.

\section{Algorithmic Complexity  of  Stochastic Systems}

The cornerstone of our argument is an abstract characterization of the
sequence  of observations  made  on a  given  system in  terms of  AIT
\cite{Solomonoff:1964,    Kolmogorov:1965,    Chaitin:1966,   Li:1997,
  Thomas:2001,  Adami:1999} -see  also  \cite{Chaitin:1975}.  The  key
quantity of such theory  is the so-called {\em Kolmogorov complexity},
which  is  a  conceptual  precursor  of statistical  entropy,  and  an
indicator of the complexity (and predictability) of a dynamical system
\cite{Nicolis:1986, Ebeling:1991,  Ebeling:1993}.  In a  nutshell, let
$\mathbf{x}$  be  a  symbolic   string  generated  by  the  successive
observations  of the  system ${\cal  S}$.  Its  Kolmogorov Complexity,
$K(\mathbf{x})$ is  defined as the  length $l(\pi^*)$-in bits-  of the
shortest program $\pi^*$ executed in  a universal computer in order to
reproduce  $\mathbf{x}$.    This  measure  has  been   often  used  in
statistical   physics   \cite{Kaspar:1987,   Zurek:1989,   Dewey:1996}
particularly in the  context of symbolic dynamics \cite{Ebeling:1991}.
In this context, $K$ is  known to be maximal for completely disordered
systems, whereas  it takes intermediate values when  some asymmetry on
the probabilities of appearance of symbols emerges.

Within   the  framework   of  statistical   physics,  a   sequence  of
observations performed  over a  given system can  be interpreted  as a
sequence  of independent,  identically  distributed random  variables,
where the specific outcomes of the observations are obtained according
to a  given probability distribution.   In mathematical terms,  such a
sequence  of  observations  defines  a {\em  stochastic  object}.   By
definition,  the   Kolmogorov  Complexity  of   a  stochastic  object,
described by  a binary  string $\mathbf{x}=x_1,..,x_m$ of  length $m$,
satisfies \cite{Grunwald:2008}:
\begin{equation}
\lim_{m\to\infty}\frac{K(\mathbf{x})}{m}=\mu\;\;\; \in (0,1].
  \label{mubin1}
\end{equation}
In other  words, the binary  representation of a stochastic  object is
{\em  linearly} compressible.   The  case where  $\mu=1$  refers to  a
completely   random   object,   and   the  string   is   called   {\em
  incompressible}.

We can generalize  the concept for non binary  strings, whose elements
belong  to a  given set  $\Sigma=\{s_1,..,s_n\}$,  being $|\Sigma|=n$.
This is  the case  of a dice,  for example,  whose set of  outcomes is
$\Sigma_{dice}=\{1,2,3,4,5,6\}$.   If the  behavior of  the  system is
governed by  the random  variable $X(n)$, accordingly,  the successive
observations   of  our   stochastic  system   define  a   sequence  of
independent,  identically $p_n$-distributed random  variables $X_1(n),
..., X_m(n)$ taking values over  the set $\Sigma$.  The so-called {\em
  noiseless    Coding}     theorem    \cite{Shannon:1948,    Ash:1990,
  Thomas:2001}, establishes that the  minimum length, (in bits) of the
string needed to code the event $s_i$, $l^*(s_i)$, satisfies
\begin{equation}
l^*(s_i)=-\log(p_n(s_i))+{\cal O}(1).
\label{Noiseless}
\end{equation}
(Throughout  the  paper, $\log\equiv\log_2$,  unless  the contrary  is
indicated).  The average minimum length will correspond to the minimum
length  of   the  code,  which  is,  by   definition,  the  Kolmogorov
complexity.  Thus we obtain \cite{Thomas:2001, Grunwald:2003}:
\begin{eqnarray}
\lim_{m\to\infty}\frac{K(X_1(n),...,X_m(n))}{m}&=&\sum_{i\leq n}p_n(s_i)l^*(s_i)\nonumber\\
&=&H(X(n))+{\cal O}(1),
\label{K(S)=H(X)}
\end{eqnarray}
being $H(X(n))$ the Shannon or statistical entropy \cite{Shannon:1948,
  Ash:1990, Thomas:2001}, namely:
\begin{equation}
H(X(n))=-\sum_{i\leq n}p_n(s_i)\log p_n(s_i). \nonumber
\end{equation}
The   complete  random   case  is   obtained  when,   $\forall  s_i\in
\Sigma\;p_n(s_i)=1/n$ leading to $l^*(s_i)=\log n +{\cal O}(1)$.  This
indicates that we  need $\approx\log n$ bits to  code any element from
$\Sigma$.   Therefore, the  length  in  bits of  the  sequence of  $m$
successive observations  will be approximately  $m \cdot \log  n.$ the
average minimum length of the code will be lower than $\log n$.  Using
our  previous result  (\ref{mubin1}) for  the binary  case, it  is not
difficult to see that:
\begin{equation}
\lim_{m\to\infty}\frac{K(X_1(n),...,X_m(n))}{m\cdot\log n}=\mu ;\;\;\mu\in (0,1].
\label{mugen}
\end{equation}
By defining $h(n)$ as the normalized entropy as:
\begin{equation}
h(n)\equiv\frac{H(X(n))}{\log n},
\end{equation}
and from eq. (\ref{K(S)=H(X)}),  we observe that eq. (\ref{mugen}) can
be rewritten as $h(n)\approx \mu;\;\;\mu\in (0,1]$.

So far we have been concerned with the algorithmic characterization of
stochastic systems  for which the  size of the configuration  space is
static.  However, we must  differentiate the properties of the systems
we want to characterize from  a standard stochastic object such as the
ones  obtained by  tossing a  dice or  a coin.   They both  generate a
bounded  number of  possible outcomes  -namely, $6$  and $2$-  with an
associated  probability, whereas  those systems  exhibiting power-laws
lack  an a  priori constraint  on  the potential  number of  available
outcomes.   These  systems are  {\em  open}  concerning  the size  -or
dimensionality- of  the configuration space.   Let $X(n)$ be  a random
variable  taking  values  on  $\Sigma$, where  $|\Sigma|=n$  and  with
associated probability distribution $p_n$,  where (without any loss of
generality) an ordering
\begin{equation}
p_n(s_1)\geq p_n(s_2)\geq...\geq p_n(s_n)
\label{OrderedProb}
\end{equation}
is assumed.  At a given time, the system satisfies eq.  (\ref{mugen}),
since  it is  a stochastic  object with  a given  number  of available
states. However, we assume that the system changes (generally growing)
maintaining its  basic statistical properties  stable \cite{Zipf:1949,
  Gabaix:1999, Axtell:2001, Rozman:1990}. Using eq. (\ref{K(S)=H(X)}),
condition (\ref{mugen}) is replaced by:
\begin{equation}
\lim_{n\to\infty}h(n)=\mu
\label{limK(X)}
\end{equation}
We  can replace  eq.  (\ref{limK(X)})  alternatively by  the following
statement: For  any $\epsilon >0$ there exists  $n\in \mathbb{N}$ such
that, for any $n'>n$:
\begin{equation}
|h(n')-\mu|<\epsilon.
\label{limheps}
\end{equation}
The main objective  of the paper is to  find the expected distribution
$p_n(s_i)$  consistent with  eq.  (\ref{limheps}).   The  case $\mu=0$
would  correspond to  systems  where, although  growing  in size,  its
complexity  (and thus, its  statistical entropy)  is bounded  or grows
sublinearily  with $\log  n$, a  case studied  in \cite{Ebeling:1991}.
Here, we are interested in the intermediate case, where
$\mu\in(0,1)$.
This  characterization would  depict systems  with some  balance among
ordering and disordering forces,  and thereby displaying a dissipation
of statistical entropy proportional  to the maximum entropy achievable
for  the system  in  equilibrium.   Therefore, we  will  refer to  the
problem  of finding  solutions for  eq.  (\ref{limK(X)})  as  the {\em
  entropy restriction problem}.
\begin{figure}
\begin{center}
\includegraphics[width= 8.5cm]{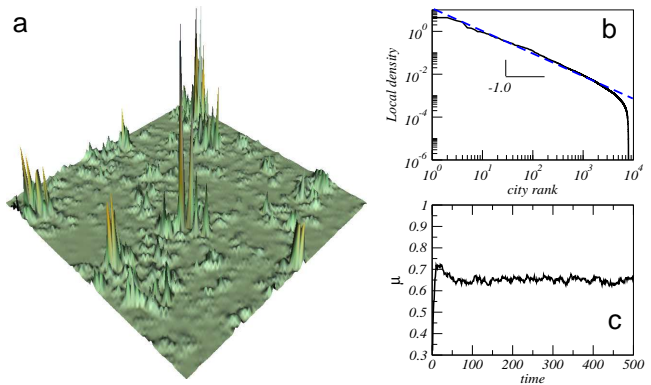}
\vspace{0.5 cm}
\caption{(Color online)  An example of the behavior  of the normalized
  entropy  for a multiplicative  stochastic process  exhibiting Zipf's
  law. Here we use the model described in \cite{Manrubia:1999} using a
  $80 \times 80$ lattice where each  node is described by a density of
  population $\rho(i,j)$. The  rules of the model are  very simple: i)
  At  every time  step, each  node loses  a fraction  $\alpha$  of its
  contents,  which is  distributed among  its four  nearest neighbors.
  ii)  At  time  $t+1$   the  local  population  is  multiplied,  with
  probability   $p$,  by   a  factor   $p^{-1}$.    Furthermore,  with
  probability   $1-p$,  the   population   of  a   node   is  set   to
  zero. Additionally, at each step  a random number $\eta$ is added to
  every node.  In this way, we  avoid falling into  an absorbing state
  $\rho=0$.  Here  we use $0<\eta  <0.01$, $\alpha =1/4$  and $p=3/4$.
  This is an  extremly simplified (and yet successful)  model of urban
  population dynamics.  A snapshot (for $t=500$) is shown in (a) where
  we  can appreciate  the  wide range  of  local densities,  following
  Zipf's law (b).  If we  plot the evolution of the normalized entropy
  $\mu$  over  time  (averaged  over  $10^2$ replicas)  we  observe  a
  convergence towards a stationary value $\mu \approx 0.65$. }
\label{Invariant}
\end{center}
\end{figure}
A computational test  for this result can be  illustrated by the model
results shown in fig.(1).  The picture shows a spatial snapshot of the
local  population densities  of  a model  of  urban growth  displaying
Zipf's  law  \cite{Manrubia:1999}.   The  normalized  entropy  evolves
towards a  stationary value $\mu  \approx 0.65$ consistently  with our
discussion.   This is  true in  spite  that this  model exhibits  wide
fluctuations due to its intermittent stochastic dynamics.

\section{Emergence of Zipf's Law in  Stochastic Systems}
As pointed  out in \cite{Harremoes:2001}, the main  difficulty we face
in this kind of equations is  that we are not dealing with an extremal
problem,  since our value  of entropy  is previously  fixed and  it is
neither  minimum  nor maximum,  in  Jaynes' sense  \cite{Jaynes:1957}.
Thus, classical variational methods,  which have been widely used with
great success in  statistical mechanics \cite{Jaynes:1957, Haken:1978,
  Kapur:1989,  Pathria:1996}, do  not apply  to our  problem -although
recently it  has been shown  that variational approaches  using Fisher
information  and physically  relevant constraints  lead  to Power-laws
whose exponent can be close to $1$ \cite{Hernando:2009}.  We also must
take into account that the  particular properties of Zipf's law create
an additional difficulty if the  studied systems display, a priori, an
unbounded  number of possible  states. Specifically,  we refer  to the
non-existence  of finite  moments  and normalization  constant in  the
thermodynamical  limit.  However,  as we  shall see,  these apparently
undesirable properties will be the key to our derivation.

\subsection{Properties of the entropies of a power law}
Let us briefly summarize the  properties of the entropies of power-law
distributed systems, which will be  used to derive the main results of
this  work  (For  details,   see  Appendix  A).  Such  properties  are
intimately  linked  with  the   behavior  of  the  {\em  Riemann  Zeta
  function}, $\zeta(\gamma)$ \cite{Abramowitz:1965}:
\begin{equation}
\zeta(\gamma)=\sum_{k=1}^{\infty}\frac{1}{k^{\gamma}}.
\label{Riemann}
\end{equation}
In  the  real   line,  this  function  is  defined   in  the  interval
$\gamma\in(1,\infty)$, displaying a singularity for $\gamma\to 1^+$.

Now,  let us  suppose  that the  system  contains $n$  states and  the
probability  to  find  the  $i$-th  most  likely  states  decay  as  a
power-law,  i.e., $p_{n}(s_i)\propto  i^{-\gamma}$.  For  the  sake of
simplicity, we  will refer to its associated  entropy as $H(n,\gamma)$
and to its normalized counterpart as $h(n,\gamma)$, i.e.:
\begin{equation}
h(n,  \gamma) =\frac{1}{\log  n}\left(
\frac{\gamma} {Z}\sum^{n}_{i=1}\frac{\log i}{i^{\gamma}}+\log Z\right).
\label{explicit}
\end{equation}
The   most   basic  properties   concern   the   global  behavior   of
$H(n,\gamma)$.  It is straightforward  to check that $H(n, \gamma)$ is
$i)$ a monotonous increasing function  on $n_t$ and $ii)$ a monotonous
decreasing function on $\gamma$.   Moreover, the normalized entropy of
Zipf's  law   of  a  system   with  $n$  states  converges   to  $1/2$
\cite{Jones:1979}, i.e.,
\begin{equation}
\lim_{n\to\infty}h(n, 1)=\frac{1}{2}.
\label{1/2}
\end{equation}
We also note that the entropy of a power law with exponent higher than
one is bounded  i.e., if $\gamma>1$ is the exponent  of our power law,
there exists a finite constant $\phi(\gamma)$ such that:
\begin{equation}
\lim_{n\to\infty}H(n,\gamma)<\phi(\gamma).
\label{bound}
\end{equation}
A key consequence of this result is that, if our (unknown) probability
distribution  is  dominated  \footnote{A probability  distribution  is
  dominated from some  $k$ by a power law  with exponent $1+\delta$ if
  $(\exists    m):(\forall    i>m)\left(\frac{p(i+1)}{p(i)}    <\left(
  \frac{i}{i+1}\right)^{1+\delta}\right)$.}   from  some  $k$ by  some
power-law with  exponent $\gamma>1+\delta$ (for any  $\delta> 0$), our
entropy will be bounded.

Furthermore,  it  can  be  shown  that the  normalized  entropy  of  a
power-law  distribution in a  system with  $n$ different  states, with
exponent $\gamma <1$, converges to $1$, i.e.,
\begin{equation}
\lim_{n\to\infty} h(n,\gamma)=1.
\label{to1}
\end{equation}
Consistently,  we  can  conclude  that, if  an  (unknown)  probability
distribution  is  not dominated  from  any $m$  by  a  power law  with
exponent lower  than $1-\delta$  (for any $\delta>0$),  the normalized
entropy of our system will converge to $1$.

Using these properties, in the following sections we proceed to derive
Zipf's  using  two complementary  approaches,  namely  1) proposing  a
power-law as the assymptotic  solution of eq. (\ref{limK(X)}) -section
IIIB- and  2) Assuming that  the entropy behaves in  a scale-invariant
way -section IIIC.

\subsection{Power Law Ansatz: Convergence of Exponents to $\gamma=1$}
\begin{figure}
\begin{center}
\includegraphics[width= 8.5cm]{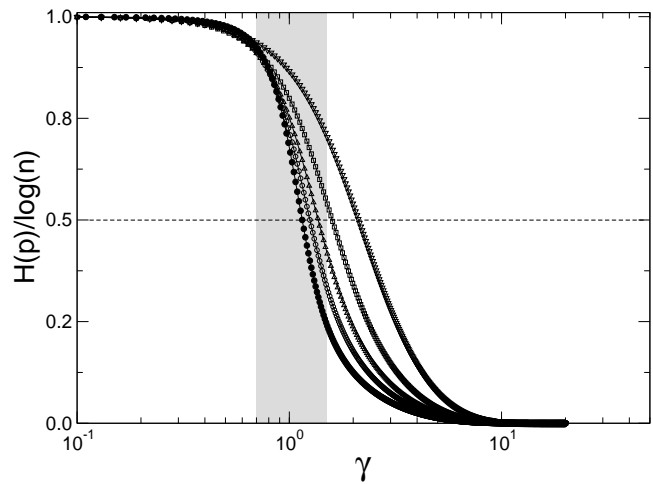}
\caption{Normalized entropies of five power-law distributed systems of
  different size as functions of the exponent.  The curves display $5$
  different sizes.  $n=500000$ black circles, $n=10000$ white circles,
  $n=10000$ up triangles, $n=1000$ squares and $n=100$ down triangles,
  respectively.   The  most   interesting  feature  of  the  numerical
  computations is the  sharp decay of the normalized  entropy when the
  values of the exponent are cllose  to $1$, which implies that a wide
  range of normalized entropies are obtained by tuning the exponent of
  the  power-law distribution around  unity.  Furthermore,  we observe
  that  the  decay  is  sharper  as  the size  of  the  system  grows,
  concentrating  an increasing  range of  relative entropies  near the
  exponent $1$ (grey area).}
\label{Fases}
\end{center}
\end{figure}
In this section  we make use of the power-law ansatz  as a solution of
our problem, i.e., we assume that  the solution is a power-law with an
arbitrary   exponent,   i.e.,   $p_n(s_i)\propto  i^{-\gamma}$.    The
objective of this section  is to demonstrate that, being $h(n,\gamma)$
as defined in eq. (\ref{explicit}), then the following limit holds:
\begin{equation}
\lim_{n\to\infty}h(n,\gamma)=\Theta(\gamma),
\end{equation}
being $\Theta(\gamma)$ the  step function, i.e., $\Theta(\gamma)=1$ if
$\gamma<1$ and  $\Theta(\gamma)=0$ if  $\gamma >1$.  It  implies that,
for  large values  of $n$,  the  whole range  of normalized  entropies
between $0$  and $1$ is  obtained from exponents  $\gamma$ arbitrarily
close to $\gamma=1$ -see fig (\ref{Fases}).

Let   us    rewrite   the   convergence    assumptions   provided   in
(\ref{limK(X)},\ref{limheps})    assuming    that   our    probability
distribution is a power law: For  any $\epsilon >0$ we can find an $n$
such that, for any $n'>n$ we have an exponent, $\gamma(n')$ such that,
\begin{equation}
|h(n', \gamma(n'))-\mu|<\epsilon,
\label{epsilon}
\end{equation}
i.e., the  sequence of normalized entropies ${\cal  H}$, associated to
system's growth, namely
\begin{equation}
{\cal H}=h(1, \gamma(1)),h(2,\gamma(2)),
..., h(k, \gamma(k)), ...,
\label{Seqh}
\end{equation}
converges  to $\mu$.   Below we  split  the problem  in two  different
scenarios.
\subsubsection{First case: $\mu<\frac{1}{2}$.}
We begin  by exploring the following scenario:
\begin{equation}
\lim_{n\to \infty}h(n,\gamma(n))=\mu\in\left(0,\frac{1}{2}\right).
\end{equation}
From equation (\ref{1/2}) we can ensure that, for large values of $n$,
$\gamma(n)>1$.   Since  we  assumed   that  the  sequence  ${\cal  H}$
converges to  $\mu$, we can  state that, for  a given $\epsilon  > 0$,
there is an arbitrary $n_1$ such that:
\begin{equation}
\mu-\epsilon<h(n_1,\gamma(n_1))<\mu+\epsilon.
\label{ineq}
\end{equation} 
We know, from the properties of the entropies of power-law distributed
systems,    that   $H(n_1,\gamma(n_1)    )<\phi(\gamma(n_1))$,   where
$\phi(\gamma(n_1))$  is  some   positive,  finite  constant  (see  eq.
(\ref{bound}) and  appendix).  Then, since  $\log x$ is  an unbounded,
increasing function of $x$, we can find $n_2>n_1$ such that
\begin{equation}
\phi(\gamma(n_1))<(\mu+\epsilon)\log n_2.
\end{equation}
Thus, since  $h(n, \gamma)$ is  a decreasing function on  $\gamma$, we
need to find $\gamma(n_2)<\gamma(n_1)$ such that
\begin{equation}
\mu-\epsilon'< h(n_2,\gamma(n_2))<\mu+\epsilon',
\end{equation}
with  $\epsilon'\leq  \epsilon$,  in  order  to  satisfy  the  entropy
restriction.   Furthermore,  since  $H(n,  1)=\frac{1}{2}\log  n+{\cal
  O}(\log(\log  n))$,  we  conclude that  $1<\gamma(n_2)<\gamma(n_1)$.
Let us  expand this process  recursively, thus generating  an infinite
decreasing sequence of exponents,
\begin{equation}  
\{\gamma(n_k)\}_{k=1}^{\infty}=\gamma(n_1),...,\gamma(n_i),...,
\label{sequence1}
\end{equation}
such  that, for  any  $\gamma(n_i)\in \{\gamma(n_k)\}_{k=1}^{\infty}$,
$\gamma(n_i)>1$.  We  notice that, for  any $\alpha>0$, we can  find a
$n_k$ such that, if $n_j>n_k$,
\begin{equation}
|\gamma(n_j)-1|<\alpha,
\end{equation}
since, for every $\gamma(n_k)$, we always find a $n_j>n_k$ such that
\begin{equation}
\phi(\gamma(n_k))<(\mu+\epsilon)\log n_j.
\end{equation}

\subsubsection{Second case: $\mu>\frac{1}{2}$.}
Let us now consider the following entropy restriction problem:
\begin{equation}
\lim_{n\to\infty}h(n, \gamma(n))=\mu \in\left(\frac{1}{2}, 1\right).
\end{equation}
From  equation   (\ref{1/2}),  we  can  ensure  that,   for  any  $n$,
$\gamma(n)<1$.  Furthermore, from  equation (\ref{to1}), we again find
a problem close to the one  solved above, since for $n_1$ large enough
and $\gamma<1$, we have:
\begin{eqnarray}
H( n_{1}+1 ,\gamma)-H(n_{1}, \gamma)>\mu(\log(n_{1}+1)-\log n_{1}).
\label{Ht1}
\end{eqnarray}
Now, since we  assumed that the sequence ${\cal  H}$ converges, we can
state that given an arbitrary step $n_1$,
\begin{equation}
\mu-\epsilon<h(n_{1},\gamma(n_{1}))<\mu+\epsilon.
\end{equation}
Since $H(n, \gamma)$ is a  decreasing function on $\gamma$, we need to
find $\gamma(n_{2})>\gamma(n_{1})$ such that:
\begin{equation}
\mu-\epsilon'< h(n_{2},\gamma(n_{2}))<\mu+\epsilon',
\end{equation}
with  $\epsilon'\leq \epsilon$,  to satisfy  the  entropy restriction.
However,     from      eq.      (\ref{1/2}),     we      know     that
$1>\gamma(n_{2})>\gamma(n_{1})$.  Proceeding as  above, we expand this
process, thus generating an  infinite increasing sequence of exponents
$\{\gamma(n_{k}\}_{k=1}^{\infty}$.  By  virtue of equation (\ref{1/2})
and  equation  (\ref{to1}), and  taking  into  account the  decreasing
behavior of  $h$ as a function  of the exponent, we  observe that, for
any $\alpha>0$, we can find a $n_k$ such that, if $n_j>n_k$,
\begin{equation}
|\gamma(n_{j})-1|<\alpha.
\end{equation}

In  summary, under the  power law  ansatz, the  only solution  for eq.
(\ref{limK(X)}), in  the limit of large systems,  is $\gamma=1$, i.e.,
Zipf's  law.   

\subsection{Scale invariance Condition}

The above  {\em power-law ansatz}  is purely mathematical, and  can be
replaced by a more physically  realistic assumption.  This leads us to
the  second strategy  to  solve our  problem,  which is  based on  the
assumption  that  the  mechanisms   responsible  for  the  growth  and
stabilization  of  the  system  do  not  depend on  the  size  of  the
configuration space,  and, thus, a  partial observation of  the system
will satisfy  also condition (\ref{limK(X)}).   We will refer  to this
assumption  as  the  {\em  scale  invariance  condition},  and  it  is
formulated as follows.  Let  $\Sigma^{(k)}\subseteq \Sigma$ be the set
of the first $k$ elements of $\Sigma$, observing a labeling consistent
with    the    ordering    of    probabilities   provided    in    eq.
(\ref{OrderedProb}) -roughly speaking,  the $k$ most probable elements
of $\Sigma$.  The random  variable which accounts for the observations
of such  $k$ elements  is notated $X(k\leq  n)$.  We observe  that, if
$X(n)$  follows  the  probability  distribution  $p_{n}$,  the  random
variable $X(k\leq n)$ obeys the following probability distribution, to
be notated $p_n^k$:
\begin{equation}
p_{n}^k(i)\equiv \mathbb{P}(s_i|i\leq k)=\left(\sum_{j\leq k}p_{n}(s_j)\right)^{-1}p_{n}(s_i).
\label{pntk}
\end{equation}
Thus,  if  $H(X(k\leq  n))$  is  the  entropy  of  $X(k\leq  n)$,  its
normalized counterpart is defined as $h(k\leq n)$:
\begin{equation}
h(k\leq n)\equiv\frac{H(X(k\leq n))}{\log k}.
\label{hn<k}
\end{equation}
We  remark that  these derivations  are valid  at the  limit  of large
systems, thereby  considering that, at every step,  $n$ is arbitrarily
greater than $k$. Furthermore, let us define $\epsilon'$ as:
\begin{equation}
\epsilon'\equiv |h(k\leq n)-\mu|+\delta,
\end{equation}
being  $\delta$   arbitrarily  small.   Then,   the  scale  invariance
assumption for the entropy states that, for any $n\geq k'\geq k$,
\begin{equation}
\left|h(k'\leq n)-\mu\right|<\epsilon'.
\label{ScaleInvh}
\end{equation}
In summary, condition (\ref{ScaleInvh}), is grounded on the assumption
that  the entropy  restriction  works at  all  levels of  observation.
Thus, the  partial probability distributions of states  we obtain must
reflect  the effect  of the  entropy restriction,  introducing  a {\em
  scale invariance}  of the normalized entropy of  the partial samples
of the system.


As we  saw in the above sections,  the decay of this  tail is strongly
constrained by the entropy restriction, since only special cases avoid
the normalized entropy to fall to  $0$ or $1$.  To study in detail how
it  constrains the  tail of  the distribution  we will  work  with the
coefficients $f_n(k, k+1)$, defined as:
\begin{equation}
f_n(i,i+1)=\frac{p_{n}(s_i)}{p_{n}(s_{i+1})},\nonumber
\end{equation}
instead  of the  raw  probability distribution,  to avoid  multiplying
factors due to normalization.  Now  we observe that, for a given, very
large $n$, our probability distributions  $p^k_n$ must be able, as $k$
increases, to unboundedly increase the  entropy of the whole system to
reach the global value $H(X(n))$,  which lies in the interval $((\mu -
\epsilon)\log n,(\mu+\epsilon)\log n)$.  Furthermore, scale invariance
condition  depicted  in eq.   (\ref{ScaleInvh})  forces  that, as  $k$
increases, {\em contributions  to the entropy never go  neither to $0$
  nor to $\log(k+1/\log k)$}, but  lie within this interval.  In other
words, the sum defined by  the entropies must diverge as $k$ increases
over a  system where $n$ is  arbitrary large, whereas  the sequence of
its normalized versions must converge to $\mu$.  The above derivations
concerning  the  convergence  properties  of  the  entropy  -see  also
Appendix  A-  clearly  state  that  those  properties  hold  if  $p_n$
satisfies, on one hand, for large $i$'s,
\begin{equation}
f_n(i, i+1)<\left(\frac{i+1}{i}\right)^{(1-\delta)},
\end{equation}
to avoid  that $h(n)\to 1$. On the  other hand, if we  want to avoid
that $h(n)\to 0$, the following inequality must hold:
\begin{equation}
f_n(i, i+1)>\left(\frac{i+1}{i}\right)^{(1+\delta)}.
\end{equation}
Therefore, the solution of our problem lies in the range defined by:
\begin{equation}
\left(\frac{i+1}{i}\right)^{(1-\delta)}>f_n(i, i+1)>\left(\frac{i+1}{i}\right)^{(1+\delta)}.
\end{equation}
From  the study  of the  entropies  of a  power law  performed in  the
previous section,  we know that  $\delta$ can be arbitrarily  small if
the size of the system is large enough. Thus:
\begin{equation}
f_n(i,i+1)=\frac{p_{n}(s_i)}{p_{n}(s_{i+1})}\approx \frac{i+1}{i}
\end{equation}
which leads us to Zipf's law as the unique asymptotic solution:
\begin{equation}
p_{n}(s_i)\propto i^{-1}.
\end{equation}

\section{Discussion}

Complex,  far  from  equilibrium  systems involve  a  tension  between
amplifying mechanisms and negative feedbacks able to buffer the impact
of fluctuations.  In this paper we have considered the consequences of
such tension  in terms  of one  of its most  well known  outcomes: the
presence of  an inverse  scaling law connecting  the size  of observed
events and  its rank.  The commonality  of Zipf's law  in both natural
and man-made  systems has been a  puzzle that attracted  for years the
attention of scientists, sociologists  and economists alike.  The fact
that such a plethora of  apparently unrelated systems display the same
statistical   pattern  points   towards  some   fundamental,  unifying
principle.

In  this  paper  we   treat  complex  systems  as  stochastic  systems
describable in  terms of  algorithmic complexity and  thus statistical
entropy. A  general result from  the algorithmic complexity  theory is
that eq. (3) holds for  stochastic systems. Taking this general result
as the starting point, we define a characterization of a wide class of
complex systems which grasps the  open nature of many complex systems,
summarized  in  eq.  (\ref{limK(X)}).  The main  achievement  of  this
equation is  that it  encodes the concepts  of growing and,  even most
important,   the  stabilization   of  complexity   properties   in  an
intermediate point  between order and disorder, a  feature observed in
many systems displaying Zipf's-like statistics.  From this equation we
derived Zipf's law as the natural outcome of systems belonging to this
class of stochastic systems.

Our development avoids the  classical procedures based on maximization
(minimization) of some  functional in order to find  the most probable
configuration of  states, since in  far from equilibrium  the ensemble
formalism,   together   with   Jaynes'   maximum   entropy   principle
\cite{Jaynes:1957} can  fail due to the  open, non-reversible behavior
of  the systems  considered here.   Thus  we do  not introduce  moment
constraints,  as  it is  usual  in  equilibrium statistical  mechanics
\cite{Pathria:1996}, but instead a constraint on the value achieved by
the normalized  entropy, no  matter the scale  we observe  the system.
Both a scaling  ansatz and a more general  scale invariance assumption
lead to Zipf's law as the unique solution for this problem. We observe
that the  finite size effects  define an interval of  exponents around
$1$,  namely  $(1-\delta,1+\delta)$, which  could  partly explain  the
variation observed  in finite, natural  systems.  However, it  is true
that  a system  satisfying eq.   (\ref{limK(X)}) does  not necessarily
exhibit  Zipf's  law.   Further  work  should  explore  in  depth  the
physically relevant  conditions leading  the evolution of  Zipf's like
systems  to remove the  mathematical assumptions  made in  this paper,
thereby  obtaining a complete  description of  them from  a completely
general, theoretical viewpoint.

\appendix
\section{Entropic Properties of Power-Law distributed systems}

Consider a system  whose behavior is described by  the random variable
$X(n)$   taking    values   on   the    set   $\Sigma   =\{s_1,...,s_n
\},\;|\Sigma|=n$,    according   to   the    probabilty   distribution
$p_n(s_i)$. The  labeling '$i$' of the  state is chosen in  such a way
that $p_n(s_1)\geq p_n(s_2)\geq...\geq p_n(s_i)\geq...\geq p_n(s_n).$

The  Shannon  entropy  of  our  system  of $n$  states,  to  be  noted
$H(X(n))$, is defined as \cite{Ash:1990}:
\begin{equation}
H(X(n))=-\sum_{k\leq n}p_n(s_k)\log p_n(s_k).
\end{equation}
The  normalized entropy  of  the  system, to  be  written, $h(n)$,  is
defined as:
\begin{equation}
h(n)\equiv \frac{H(X(n))}{\log n}.
\end{equation}
We  will  work  with  power-law  distributions,  by  which  $p_n(s_i)=
\frac{1}{Z}i^{-\gamma}$ where $n$ is  the number of available states,
and $Z$ the  normalization constant, which depends on  the size of the
system, $n$.  Let  us rewrite the function $H(X(n))$  as a function of
the exponent and the size of $\Sigma$, $H(n,\gamma)$.  Consistently,
\begin{equation}
h(n,\gamma)\equiv\frac{H(n,\gamma)}{\log n}.
\end{equation}

This appendix is  devoted to derive five properties  of the entropy of
power-law distributed systems.

{\bf 1}.   {\em $H(n,\gamma)$  is a continuous,  monotonous decreasing
  function with respect to $\gamma$ in the range $(0,\infty)$. }

Indeed, the dominant term of its derivative is:
\begin{equation}
\frac{\partial    H(n, \gamma)}{\partial\gamma}   \sim    -\sum_{i\leq
  n}\frac{(\log i)^2}{i^{\gamma}}<0.
\label{decrec}
\end{equation}

{\bf 2}.{\em  The entropy of  a power-law is a  monotonous, increasing
  function on the size of the system}\footnote{The reader could object
  that this section is  unnecessary, since the axiomatic derivation of
  the uncertainty  function (which we  call entropy) assumes  that the
  entropy increases with the size of the system. However, the explicit
  statement of this  axiom corresponds to the special  case of uniform
  probabilities \cite{Ash:1990}.  Specifically, the axiom states that,
  {\em if we have two systems $A, B$ such that $A$ contains $n$ states
    $a_1,...,a_n$  and $B$  contains $n+1$  states, $b_1,...,b_{n+1}$,
    then, if  $(\forall i\leq n) p(a_i)=1/n$ and  $(\forall i\leq n+1)
    p(b_i)=1/(n+1)$}
    $H(A)<H(B)$.
Thus,  if we  are  not dealing  with  this special  case,  we need  to
explicitly demonstrate that it holds  for our purposes.}.  

We want to show that $H(n,\gamma)$ is a monotonous increasing function
on  $n$.   In  order to  prove  it,  we  must compute  the  difference
$H({n+1}, \gamma)-H(n, \gamma)$.  For simplicity, let us define:
\begin{equation}
S_n\equiv \sum_{k\leq n}\frac{1}{k^{\gamma}}.
\end{equation} 
Using the trivial inequality:
\begin{equation}
\log\left(S_n+\frac{1}{(1+n)^{\gamma}}\right)>\log(S_n),
\end{equation}
we can state that:
\begin{widetext}
\begin{eqnarray}
H({n+1}, \gamma)-H(n, \gamma)&=&
 \frac{\gamma}{S_n+\frac{1}{(1+n)^{\gamma}}}\sum_{k\leq n+1}\frac{\log k}{k^{\gamma}}+\log\left(S_n+\frac{1}{(1+n)^{\gamma}}\right)-\frac{\gamma}{S_n}\sum_{k\leq n}\frac{\log k}{k^{\gamma}}+\log\left(S_n \right)\nonumber\\
&>& \gamma\sum_{k\leq n}\frac{\log k}{k^{\gamma}}\left(\frac{1}{S_n +\frac{1}{(n+1)^{\gamma}}}-\frac{1}{S_n}\right)+\gamma\frac{\log(n+1)}{S_n+\frac{1}{(n+1)^{\gamma}}}\nonumber\\
&=&\frac{\gamma}{S_n^2(n+1)^{\gamma}+S_n}\left(S_n(n+1)^{\gamma}\log(n+1)-\sum_{k\leq n}\frac{\log k}{k^{\gamma}}\right)\nonumber\\
&>&0.\nonumber
\end{eqnarray}
\end{widetext} 
Finally, it is easy to check that the following properties also hold:
\begin{eqnarray}
\lim_{\gamma\to \infty}H(n,\gamma)&=&0\label{lim0},\\
\lim_{\gamma\to 0} H(n,\gamma)&=&\log n.\label{limn}
\end{eqnarray}


{\bf 3}.  {\em The normalized entropy  of Zipf's law of  a system with
  $n$ states ($p_n(s_i) \propto i^{-1}$) converges to $1/2$:}

We want to show that the sequence
\begin{equation}
{\cal H}=\{h(k,1) \}_{k=1}^{\infty}=h(1,1), h(2,1), ..., h(k,1),...
\end{equation}
converges  to $\frac{1}{2}$.   Let us  suppose  that ${\cal  H}$ is  a
sequence satisfying  the above requirements.  Then, the  entropy for a
given $n$ can be approached by \cite{Jones:1979}:
\begin{equation}
H(n,1)=\frac{1}{2}\log n+{\cal O}(\log(\log n)).
\end{equation}
Thus, if $h(n,1)=H(n,1)/\log n$, let us define $\epsilon(n)$ like:
\begin{equation}
\epsilon(n)\equiv\left|h(n,1)-\frac{1}{2}\right|=\left|\frac{{\cal O}(\log(\log n))}{\log n}\right|.
\end{equation}
Clearly,   $\epsilon(n)$   is  strictly   decreasing   on  $n$,   and,
furthermore,
\begin{equation}
\lim_{n\to\infty}\epsilon(n)=0.
\end{equation}

{\bf 4}. {\em The Entropy of  a power law with exponent higher  than $1$ is bounded.}

Here we demonstrate  that the entropy  of a power  law with
exponent  higher  than  $1$  is  bounded\footnote{This  derivation  is
  equivalent to  the one found in \cite{Harremoes:2001},  Theorem 8.2. In
  this   theorem,  the   authors  demonstrate   that   every  infinite
  distribution  with  infinite  entropy  is  {\em  hyperbolic},  which
  implies that the  distribution is not dominated by  a power law with
  an exponent higher than $1$.}.  Specifically, we assume there exists
a pair of positive constants $Z, \delta$, such that:
\begin{equation}
p_n(i)=\frac{1}{Z}i^{-(1+\delta)}
\end{equation}
Then,  the sequence of  ${\cal H}=\{h(k,  1+\delta) \}_{k=1}^{\infty}$
converges to $0$.  Indeed, let us first note that:
\begin{equation}
\lim_{n\to \infty}p_n(s_i)=\frac{1}{\zeta(1+\delta)}i^{-(1+\delta)},
\end{equation}
where 
\begin{equation}
\zeta(1+\delta)\equiv\sum_k^{\infty}\frac{1}{k^{1+\delta}}
\end{equation}
is the Riemann  zeta-function \cite{Abramowitz:1965}.  The function is
defined  by an  infinite sum  which converges,  in the  real  line, if
$\delta>0$, i.e.:
\begin{equation}
\sum_k^{\infty}\frac{1}{k^{1+\delta}}<\infty.
\end{equation}
otherwise, the  sum diverges.  Furthermore,  it is also true  that the
above condition also holds for the following series:
\begin{equation}
\sum_k^{\infty}\frac{\log k}{k^{1+\delta}}.
\end{equation}
Indeed, note that, given an arbitrary $\delta>0$ there exists a finite
number $i^*$ such that:
\begin{equation}
i^*\equiv \min\left\{ i: \left(\delta-\frac{\log(\log i)}{\log i}\right)>0\right\}
\end{equation}
and, if we define the following exponent, $\beta(i^*)$:
\begin{equation}
\beta(i^*)\equiv 1+\delta-\frac{\log(\log i^*)}{\log i^*},
\end{equation}
there exists a finite constant, $\Psi(\delta)$, defined as:
\begin{eqnarray}
\Psi(\delta)\equiv \sum_{i< i^*}\left(\frac{\log i}{i^{1+\delta}}-\frac{1}{i^{\beta(i^*)}}\right)+\zeta(\beta(i^*)),
\end{eqnarray}
such that:
\begin{eqnarray}
\sum_k^{\infty}\frac{\log k}{k^{1+\delta}}<\Psi(\delta).
\end{eqnarray}

With  the  above properties,  it  is clear  that,  if  there exists  a
constant $\phi(1+\delta)<\infty$ such that:
\begin{equation}
\lim_{n\to\infty}H(n, 1+\delta)<\phi(1+\delta),
\end{equation}
then, the  entropy of  a power  law with exponent  higher than  $1$ is
bounded. As we  shall see, it is straightforward  by checking directly
the behavior of $H(n, 1+\delta)$:
\begin{eqnarray}
\lim_{n\to \infty}H(n, 1+\delta)
=\frac{1+\delta}{\zeta(1+\delta)}\sum^{\infty}_{i=1}\frac{\log i}{i^{1+\delta}}+\log(\zeta(1+\delta)).\nonumber
\label{lim}
\end{eqnarray}
Since $H(n,\gamma)$ is an increasing function on $n$, and
\begin{eqnarray}
\frac{1+\delta}{\zeta(1+\delta)}\sum^{\infty}_{i=1}\frac{\log i}{i^{1+\delta}}+\log(\zeta(1+\delta))<\infty,
\end{eqnarray}
we can define a constant $\phi(1+\delta)$,
\begin{equation}
\phi(1+\delta)\equiv \lim_{n\to \infty}H(n, 1+\delta)+\epsilon
\end{equation}
(where $\epsilon$ is any positive, finite constant). Clearly,
\begin{equation}
H(n, 1+\delta)< \phi(1+\delta).
\label{philess}
\end{equation}
Thus, 
\begin{eqnarray}
\lim_{n\to \infty} h(n,1+\delta)&=&\lim_{n\to \infty}\frac{H(n,1+\delta)}{\log n}\nonumber\\
& \leq &\lim_{n\to \infty}\frac{\phi(1+\delta)}{\log n}\nonumber\\
&=&0. \nonumber
\end{eqnarray}

{\bf  Consequence} {\em  If an  unknown probability  distribution is
  dominated from some $k$ by  some power-law with exponent higher than
  $1+\delta$, our entropy will be bounded.}

{\bf  Consequence}  {\em If  an  unknown  probability distribution  is
  dominated from some $k$ by  some power-law with exponent higher than
  $1+\delta$, our normalized entropy will tend to $0$.}


{\bf 5}.  {\em The normalized entropy of a power-law distribution in a
  system with $n$ different states, $p_n$ with exponent lower than $1$
  converges to $1$}.

Let us  suppose that we  have the following  probability distribution,
with $ 0<\delta<1$:
\begin{equation}
p_{n}(s_i)= \frac{1}{Z}i^{-(1-\delta)}.
\end{equation}
Note that \cite{Jones:1979}:
\begin{equation}
\sum_{k\leq n}\frac{1}{k^{1-\delta}}=\int_1^{n}\frac{1}{x^{1-\delta}}+{\cal O}(1)\approx\frac{n^\delta}{\delta}.
\end{equation}
Applying directly the definition of entropy,
\begin{equation}
H(n, 1-\delta)=\frac{\delta(1-\delta)}{n^{\delta}}\sum_{k\leq n}\frac{\log k}{k^{1-\delta}}+\delta\log n-\log\delta.
\end{equation}
If we compute the limit of $h(n,1-\delta)$:
\begin{eqnarray}
\lim_{n\to \infty}h(n,1-\delta)&=&\lim_{n\to\infty}\left(\frac{\delta(1-\delta)}{\log n\cdot n^{\delta}}\sum_{k\leq n}\frac{\log k}{k^{1-\delta}}+\delta\right)\nonumber\\
&=&\lim_{n\to\infty}\frac{1-\delta}{\log n}\left(\log n -\frac{1}{\delta}\right)+\delta\nonumber\\
&=&1-\delta+\delta\nonumber\\
&=&1.\nonumber
\end{eqnarray}

{\bf Consequence}  {\em If  our (unknown) probability  distribution is
  not dominated from some $k$ by a power law with exponent higher than
  $1-\delta$, our normalized entropy will converge to $1$.}


\begin{acknowledgments}
We thank D.   Jou, S.  Manrubia, J. Fortuny and  our colleagues at the
Complex Systems  Lab for their  useful comments.  We  also acknowledge
the helpful comments provided by two anonymous referees. This work has
been founded  by the  McDonnell Foundation (BCM)  and by the  Santa Fe
Institute (RS).
\end{acknowledgments}

\end{document}